\documentclass[prb,twocolumn,floatfix,showpacs]{bla}
\usepackage{graphicx,color}
\usepackage{amsmath,amssymb,bm}

\newcommand{\mc}{\mathcal}
\newcommand{\la}{\Lambda}
\newcommand{\tn}{\textnormal}
\newcommand{\cprb}[3]{Phys.~Rev.~B {\bf #1}, #2 (#3)}
\newcommand{\cprl}[3]{Phys.~Rev.~Lett.~{\bf #1}, #2 (#3)}
\newcommand{\cnjp}[3]{New J.~Phys.~{\bf #1}, #2 (#3)}
\newcommand{\cjp}[3]{J.~Phys.: Condensed Matter {\bf #1}, #2 (#3)}
\newcommand{\cnature}[3]{Nature {\bf #1}, #2 (#3)}

\definecolor{darkred}{rgb}{0.90,0,0}
\definecolor{darkgreen}{rgb}{0,0.60,.2}
\definecolor{darkblue}{rgb}{0,0,1}
\definecolor{grey}{cmyk}{0,0,0,0.25}
\definecolor{orange}{cmyk}{0,0.6,0.8,0}

\begin{document}
\title{\boldmath Supercurrent through a multi-level quantum dot close to singlet-triplet degeneracy}
\author{C.\ Karrasch$^{1,2}$}
\author{S.\ Andergassen$^2$}
\author{V.\ Meden$^2$}

\affiliation{$^1$Department of Physics, University of California, Berkeley, California 95720, USA}

\affiliation{$^2$Institut f\"ur Theorie der Statistischen Physik and JARA -- Fundamentals of Future Information Technology, RWTH Aachen University, 52056 Aachen, Germany}

\begin{abstract}
We investigate two serially-aligned quantum dots in the molecular regime of large tunnel couplings $t$. A Zeeman field $B$ is used to tune the energy difference of singlet and triplet spin configurations. Attaching this geometry to BCS source and drain leads with gap $\Delta$ and phase difference $\phi$ gives rise to an equilibrium supercurrent $J$. To compute $J$ in presence of Coulomb interactions $U$ between the dot electrons, we employ the functional renormalization group (FRG). For $B\approx t$ -- where the singlet and (one out of a) triplet spin states are equal in energy -- the current exhibits characteristics of a $0$-$\pi$ transition similar to a single impurity. Its magnitude in the $\pi$ phase, however, jumps discontinuously at $B=t$, being smaller on the triplet side. Exploiting the flexibility of the FRG, we demonstrate that this effect is generic and calculate $J$ for realistic experimental parameters $\Delta$, $U$, and gate voltages $\epsilon$. To obtain a more thorough understanding of the discontinuity, we analytically treat the limit $\Delta=\infty$ where one can access the exact many-particle states. Finally, carrying out perturbation theory in the dot-lead couplings substantiates the intuitive picture that Cooper pair tunneling is favored by a singlet spin configuration while inhibited by a triplet one.

\end{abstract}

\maketitle

\section{Introduction}
\label{sec:intro}

Many intriguing phenomena in condensed matter physics are to be credited to many-body interactions. Two prominent examples are the Kondo effect as well as superconductivity. If a magnetic impurity is placed inside a superconducting metal, the low-energy behavior is governed by a competition of both. Depending on the ratio of the Kondo temperature $T_K$ and the BCS gap $\Delta$, Cooper pairs are broken up in order to screen the local spin ($T_K\gg\Delta$, `$0$ phase'); or bulk superconductivity prevails and the impurity harbors a free magnetic moment ($T_K\ll\Delta$, `$\pi$ phase'). At zero temperature $T=0$ and Zeeman field $B=0$, a first order quantum phase transition from a non-degenerate to a twofold degenerate ground state takes place as the ratio $\Delta/T_K$ increases. This overall picture was essentially established decades ago;\cite{oldwork1,oldwork2,oldwork3} yet, the advent of nanotechnology now renders it feasible to fabricate quantum dots (`magnetic impurities') attached to BCS source and drain leads. The various microscopic parameters of these systems in addition to $T_K$ and $\Delta$, such as a gate voltage $\epsilon$ or left-right asymmetry in the dot-lead couplings $\Gamma_{L,R}$, can be accurately determined (and possibly varied) in the experiments. This allows to study the physics -- and thus correlations -- in a very controlled way.\cite{exp1,exp2,exp3,exp4,exp5,exp6,exp7,exp8,expnovotny,expdaenen,helene}

Consequently, a renewed interest in the interplay of superconductivity and the Kondo effect and how it manifests in the equilibrium supercurrent $J$ (that generally flows between source and drain in presence of a finite phase difference $\phi$) for the particular quantum dot setup was triggered. Most theoretical studies focused on a single Anderson impurity coupled to BCS leads as a minimal (yet challenging) model,\cite{gm,ra,vecino,oguri1,florens,choi,novotny,se,oguri2,bcspaper,qmc} and eventually both the boundary of the $0$-$\pi$ phase transition as well as $J$ were calculated using reliable many-particle methods.\cite{oguri1,florens,choi,se,oguri2,bcspaper,qmc} By adjusting all microscopic parameters to the ones associated with an experimental device, the critical supercurrent and hallmarks of the $0$-$\pi$ transition were indeed calculated in qualitative agreement with the measured data.\cite{helene} The functional renormalization group (FRG), which allows to flexibly treat arbitrary gate voltages $\epsilon$, gaps $\Delta$, or asymmetries in the source-drain couplings $\Gamma_{L,R}$, provided a convenient tool to achieve this goal.

A realistic experimental quantum dot setup certainly features more than one single-particle level.\cite{absfootnote} However, if the energy spacing $\delta$ between subsequent levels is large compared to their broadening $\Gamma$, it is reasonable to assume that only the one closest to the Fermi energy determines the low-energy behavior. The experiments mentioned above show a series of well-separated resonances and can apparently be described by such a single-level scenario; but the progress in measurement and device fabrication technologies makes it likely that multi-level quantum dots (in the sense that $\delta\lesssim\Gamma$) or more complex geometries where several dots are coupled will be subject of upcoming studies. This motivates us to address the intriguing physics of such multi-level setups with BCS leads theoretically. To this end, one needs to resort to multi-impurity Anderson models,\cite{bruder,egger1,egger2} which over the last years were widely investigated for normal leads.

In this paper, we study a geometry of two single-level dots aligned in series and attached to BCS leads. We focus on the parameter regime where the hopping $t$ between the dots is strong compared to the lead coupling $\sim\Gamma$ and explicitly demonstrate that the physics -- e.g., a $0$-$\pi$ phase transition governing the Josephson current -- can indeed be explained by single-level scenarios associated with either the bonding or anti-bonding molecular states. One might naively hope that the same simple picture emerges in presence of a Zeeman field $B\approx t$ where the single-particle (interaction $U=0$) energies of bonding spin-up and anti-bonding spin-down states are equal; or rephrased, where the smallest two-particle eigenenergies of an isolated dot (with $U>0$), which are a spin singlet as well as one out of a triplet, are almost degenerate. However, this turns out to be only partially true: Whereas for $B\approx t$ and half filling various characteristics -- e.g., the very idea of a level crossing phase transition as well as the corresponding lineshapes and parameter dependencies of the current -- are just as they are in a single-level case, the magnitude of $J$ changes discontinuously at $B=t$ in one of the phases. This indicates an \textit{additional} first order (singlet-triplet) quantum phase transition. Treating the problem within a functional renormalization group approach, we illustrate that this scenario is generic and compute $J$ as a function of the gate voltage for experimentally relevant values of $U$, $\Delta$, and $\Gamma$. To gain a more thorough understanding of the singlet-triplet transition (which we expect to be a distinct feature of any more complex quantum dot geometry close a singlet-triplet degeneracy), we supplement our FRG analysis by an exact solution of the limit $\Delta=\infty$ as well as with perturbation theory in the dot-lead couplings.

The serial quantum dot geometry with BCS leads was recently studied in Ref.~\onlinecite{rok} using the numerical renormalization group and considering the limit $t\ll\Gamma$ opposite to our one. Furthermore, Ref.~\onlinecite{rok} only focused on the limit $U=\infty$, $B=0$, and fixed $\epsilon$. The idea of a singlet-triplet transition was discussed in Ref.~\onlinecite{lee} for a general two-level geometry. However, a detailed investigation of this phenomenon, of how it manifests in $J(\epsilon)$, and of how it can be interpreted intuitively in the context of a serial quantum dot for experimentally motivated values of $U$, $\Delta$, and $\Gamma$ is still missing. This work aims at filling this gap.

We organize our paper as follows. The model is introduced in Sec.~\ref{sec:model}. The details of the functional RG approach, the treatment of the limit $\Delta=\infty$ as well as perturbation theory in $\Gamma$ are presented in Secs.~\ref{sec:frg}, \ref{sec:atom}, and \ref{sec:pt}, respectively. We present and discuss our results in Sec.~\ref{sec:results} and conclude with a brief summary. Finite-temperature data obtained at $\Delta=\infty$ is shown in the Appendix.

\begin{figure}[t]
   \includegraphics[width=0.9\linewidth,clip]{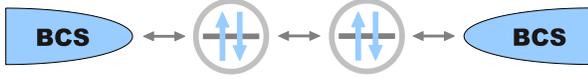}
   \caption{The multi-level quantum dot geometry subject to this paper.}
   \label{fig:model}
\end{figure}

\section{Model \& Method}

\subsection{Model}
\label{sec:model}

Our multi-level quantum dot Josephson geometry is depicted in Fig.~\ref{fig:model}. The BCS source and drain (i.e., left and right) leads are modeled by the Hamiltonian (in momentum space and using standard second-quantized notation)
\begin{equation}
H_\tn{lead}^{s=L,R}=
\sum_{k\sigma}\epsilon_{k}c^\dagger_{sk\sigma}c_{sk\sigma}^{\phantom{\dagger}} 
- \Delta\sum_k\left(e^{i\phi_s}c^\dagger_{sk\uparrow}c^\dagger_{s-k\downarrow} + \tn{H.c.}\right)~,
\end{equation}
with $\Delta$ and $\phi_{L,R}=\pm\phi/2$ being the BCS gap and phase, respectively. The two Anderson impurities coupled by a hopping element $t$ are described by
\begin{equation}\begin{split}
H_\tn{dot}  &= \sum_{i=1,2}\left[\left(\tilde\epsilon+B\right)n_{i\uparrow}+\left(\tilde\epsilon-B\right)n_{i\downarrow}\right] \\
& + U \sum_{i\sigma\neq i'\sigma'} n_{i\sigma} n_{i'\sigma'}
 - t \sum_\sigma\left( d_{1\sigma}^\dagger d_{2\sigma}^{\phantom{\dagger}} + \tn{H.c.}\right)~,
\end{split}\end{equation}
where $n_{i\sigma}=d_{i\sigma}^\dagger d_{i\sigma}^{\phantom{\dagger}}$, and $i=1,2$ labels a Wannier basis. The gate voltage $\tilde\epsilon=\epsilon-3U/2$ is shifted such that $\epsilon=0$ corresponds to the point of particle-hole symmetry at zero Zeeman field $B=0$.\cite{directionb} Moreover, $U$ denotes the strength of the Coulomb interaction. We refrain from introducing a level splitting and take local as well as nearest-neighbor interactions to be equal -- yet not for fundamental reasons, but merely because it does not alter our main results. If necessary, both can be incorporated with the functional RG approach. For the same reason, the dot is assumed to couple equally to the left and right lead (of size $N\to\infty$; $\tau$ is the local hopping strength):
\begin{equation}\begin{split}
H_\tn{T} & = \sum_{s=L,R}\sum_{\sigma} \left(H_{\tn{T},+}^{s\sigma} + H_{\tn{T},-}^{s\sigma}\right)~, \\
H_{\tn{T},-}^{L\sigma} & = \big(H_{\tn{T},+}^{L\sigma}\big)^\dagger = -\frac{\tau}{\sqrt{N}} \sum_{k} c_{Lk\sigma}^\dagger d_{1\sigma}^{\phantom{\dagger}}~, \\
H_{\tn{T},-}^{R\sigma} & = \big(H_{\tn{T},+}^{R\sigma}\big)^\dagger = -\frac{\tau}{\sqrt{N}} \sum_{k} c_{Rk\sigma}^\dagger d_{2\sigma}^{\phantom{\dagger}}~.
\end{split}\end{equation}

A key ingredient to the methods discussed in the next Section is the dot Matsubara Green function $\mc G_0(i\omega)$ associated with the total Hamiltonian
\begin{equation}
H = H^L_\tn{lead} + H^R_\tn{lead} + H_\tn{dot}+H_\tn{T}
\end{equation}
in absence of Coulomb interactions ($U=0$). It can be derived straightforwardly by expressing $H$ in terms of Nambu spinors and subsequently employing equation of motion techniques.\cite{bcspaper} The result reads
\begin{equation}\label{eq:gf}\begin{split}
&\mc G_0(i\omega)=  \\
&~~\begin{pmatrix}
i\tilde\omega -\epsilon-B & \tilde\Delta & t & 0 \\
\tilde\Delta^* & i\tilde\omega +\epsilon-B & 0 & -t \\
t & 0 & i\tilde\omega -\epsilon-B & \tilde\Delta^* \\
0 & -t & \tilde\Delta & i\tilde\omega +\epsilon-B
\end{pmatrix}^{-1}~,
\end{split}\end{equation}
where we have defined
\begin{equation}
\tilde\omega = \omega\left(1+\frac{\Gamma}{\sqrt{\omega^2+\Delta^2}}\right)~,~~
\tilde\Delta = \frac{\Gamma\Delta}{\sqrt{\omega^2+\Delta^2}}e^{i\phi/2}~.
\end{equation}
Assuming a constant local density of states $\rho$ in the leads in absence of superconductivity (wide-band limit) gives rise to an energy-independent hybridization
\begin{equation}
\Gamma=\pi\tau^2\rho ~.
\end{equation}

\subsection{Functional renormalization group}
\label{sec:frg}

A recently developed approach to compute low-energy transport properties of correlated quantum impurity systems is provided by the functional renormalization group (FRG).\cite{frgrevmod} The latter reformulates a given many-particle problem in terms of an infinite set of coupled flow equations for imaginary frequency (or time) single-particle irreducible vertex functions with an infrared cutoff as the flow parameter. Truncation of this hierarchy renders the FRG approximate w.r.t.~the two-particle interaction and can hence \textit{a priori} be justified only in the limit of small $U$. The most simple truncation scheme keeps track of the self-energy $\Sigma^\la$ as well as of an effective Coulomb interaction $U^\la$ (i.e., the static part of the two-particle vertex). It yields flow equations for effective system parameters and can thus be regarded as an `RG enhanced Hartree-Fock' approach -- that does, however, not suffer from any mean field artifacts. Application of this scheme to various quantum dot geometries in equilibrium turned out to give accurate results for the zero-temperature linear conductance even at fairly large Coulomb interactions.\cite{dotsystems} Finite temperatures or non-linear finite-bias transport can be addressed by employing a more elaborate truncation\cite{ralf,frequenzen,severinsiam,fermipaper} or by a generalization to Keldysh space,\cite{severinsiam,irlmfrg,irlmprl} respectively.

In the context of superconducting leads, the Hartree-Fock-like FRG approximation was shown to describe the $T=0$ physics -- such as the boundary between the $0$ and $\pi$ phases or the Josephson current -- of the single impurity case quantitatively at small to intermediate values of the Coulomb interaction and at least qualitatively at large $U$, again without being plagued by artifacts such as the breaking of spin symmetry;\cite{bcspaper} it is thus reasonable that the same holds for the present problem. We will \textit{a posteriori} substantiate this by comparing with exact results obtained at $\Delta=\infty$.

The zero-temperature FRG flow equations for the multi-level quantum dot geometry can be obtained by a straightforward generalization of the single-level case.\cite{bcspaper} They read
\begin{equation}\label{eq:flowse}
\partial_\la\Sigma^\la_{uj} = - \frac{1}{2\pi}\sum_{kl}U^\la_{ukjl}\left[\tilde{\mc G}^\la_{lk}(i\la)
+\tilde{\mc G}^\la_{lk}(-i\la)\right]
\end{equation}
for the self-energy as well as
\begin{widetext}
\begin{equation}\label{eq:flowu}\begin{split}
\partial_\la U^\la_{ujkl} = \frac{1}{4\pi}\sum_{mnpq}\sum_{\omega=\pm\la}\Big\{
 \tilde{\mc G}^\la_{mn}(i\omega) \tilde{\mc G}^\la_{pq}(-i\omega)
U^\la_{ujmp}U^\la_{nqkl}
+ 2 \tilde{\mc G}^\la_{mn}(i\omega) \tilde{\mc G}^\la_{pq}(i\omega) \big[U^\la_{jqkm}U^\la_{nupl} - U^\la_{uqkm}U^\la_{njpl}\big]\Big\} 
\end{split}\end{equation} 
\end{widetext}
for the effective Coulomb interaction. The subscripts $j=1\ldots4$ label the four Nambu indices associated with the electron spin and the two dot orbital degrees of freedom; the Green function at scale $\la$ is introduced as
\begin{equation}
\tilde{\mc G}^{\la}(i\omega)=[\mc G_0(i\omega)^{-1}-\Sigma^\la]^{-1}~.
\end{equation}
Supplemented by the initial conditions\cite{bcspaper} $\Sigma^{\la\to\infty}=0$ as well as $U^{\la\to\infty}_{1212}=U^{\la\to\infty}_{2323}=-U^{\la\to\infty}_{1313}=U^{\la\to\infty}_{1414}=U^{\la\to\infty}_{2323}=-U^{\la\to\infty}_{2424}=-U$ (and all antisymmetric permutations), the ordinary coupled differential equations (\ref{eq:flowse}) and (\ref{eq:flowu}) can be easily solved down to $\la=0$ using standard numerical routines.

The Josephson current is defined as $J=-i\langle[H,N_R]\rangle$, with $N_{R}$ being the particle number operator of the right lead; we take $\hbar=1$ as well as the electron charge $e=1$. It can be expressed analytically in terms of the exact interacting Green function $\mc G(i\omega)$:\cite{bcspaper}
\begin{equation}\label{eq:j}
J = -2\,\tn{Im}\sum_\sigma\langle H_{\tn{T},-}^{R\sigma}\rangle = -\tn{Im}\int\frac{2\Gamma\Delta e^{-i\phi/2}}{\pi\sqrt{\omega^2+\Delta^2}}\,\mc G_{43}(i\omega)\,d\omega~,
\end{equation}
for which we eventually plug in the FRG approximation $\mc G = \tilde{\mc G}^{\la=0}$.

\subsection{Exact solution at $\Delta=\infty$}
\label{sec:atom}
The limit of large BCS gaps allows for an exact solution.\cite{oguri2,bcspaper,koenig2,koenig} Namely, the non-interacting Green function of Eq.~(\ref{eq:gf}) becomes frequency-independent at $\Delta=\infty$ except for the trivial contribution $i\omega$, and one can therefore equivalently solve the Hamiltonian
\begin{equation}\label{eq:hatom}
H_\tn{eff} = H_\tn{dot} + \Gamma \left(e^{-i\phi/2} d_{1\uparrow}d_{1\downarrow} + e^{i\phi/2} d_{2\uparrow}d_{2\downarrow} + \tn{H.c.}\right)~.
\end{equation}
This can be achieved straightforwardly by again introducing Nambu spinors, decomposing the 16-dimensional Hilbert space that underlies $H_\tn{eff}$ into sectors with different Nambu particle numbers and subsequently computing all eigenvalues $E_i$ analytically (the corresponding expressions are lengthy and will not be presented here). Moreover, it will prove helpful -- particularly when substantiating our interpretation of a singlet-triplet transition -- to characterize the associated eigenstates in terms of the square $\vec s^{\,2}$ and the $z$-component $s_z$ of the total spin:
\begin{equation}
\vec s = \frac{1}{2}\sum_{i=1,2}\sum_{\sigma\sigma'} d_{i\sigma}^\dagger\vec \sigma_{\sigma\sigma'} d_{i\sigma'}^{\phantom{\dagger}}~,
\end{equation}
where $\vec\sigma=(\sigma_x,\sigma_y,\sigma_z)^T$ is the usual vector containing the Pauli matrices $\sigma_{x,y,z}$. It is easy to show that $\vec s^{\,2}$ and $s_z$ commute with each other as well as with $H_\tn{eff}$; moreover, one should note that $2s_z+2$ is just the Nambu particle number. Having calculated the set of eigenvalues $E_i$, the supercurrent can eventually be obtained from the phase derivative of the free energy:\cite{bcspaper}
\begin{equation}\label{eq:jatom}
J = 2\partial_\phi F~,~~~F=-T\ln\sum_i e^{-E_i/T}~.
\end{equation}

\begin{figure*}[t]
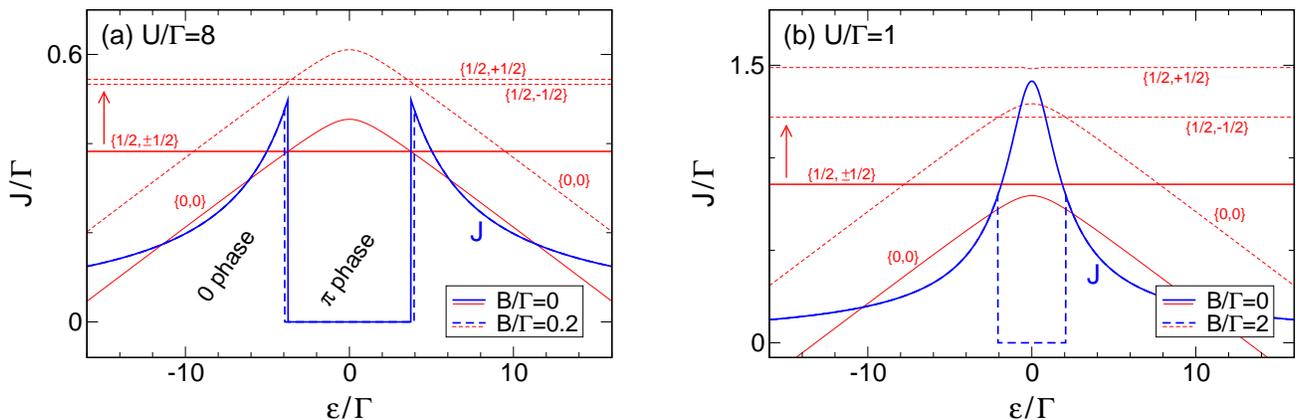

\centering
\includegraphics[width=0.45\linewidth,clip]{single1.eps}\hspace*{0.05\linewidth}
\includegraphics[width=0.45\linewidth,clip]{single2.eps}
\caption {(Color online) Gate voltage $\epsilon$ dependence of the Josephson current (thick blue lines) as well as of the three lowest many-particle energies (thin red lines) for a \textit{single} Anderson impurity coupled to BCS leads in the large-gap limit $\Delta=\infty$ at phase difference $\phi/\pi=0.5$, different Coulomb interactions $U$ and Zeeman fields $B$. The many-particle energies are shown in arbitrary units; those for finite $B$ were shifted upwards for clarity as indicated by the arrow. The corresponding eigenstates are characterized by spin quantum numbers $\{s,s_z\}$.  }
\label{fig:single}
\end{figure*}

\subsection{Perturbation theory in $\Gamma$}
\label{sec:pt}

The functional renormalization group approximation is strictly controlled in the limit of small Coulomb interactions. A complementary approach (yet not RG-based) is to carry out an expansion w.r.t.~the tunnel Hamiltonian $H_\tn{T}$, which can be justified if both the gap $\Delta$ as well as the distance between the many-particle eigenstates of $H_\tn{dot}$ are large compared to $\Gamma$.\cite{gm,novotny} The resulting perturbation theory expression for the Josephson current can be interpreted in terms of successive tunnel events. Such intuitive picture will turn out to be helpful for understanding the FRG results for the problem at hand.

The current of Eq.~(\ref{eq:j}) can be easily expanded w.r.t.~$H_\tn{T}$ (see Ref.~\onlinecite{novotny} for an instructive and detailed treatment of the single-level case). The first non-vanishing term reads
\begin{equation}\label{eq:pt1}\begin{split}
J & = \frac{1}{3}\,\tn{Im}\sum_\sigma \int_0^{\beta}d\tau_1d\tau_2d\tau_3\\ & \hspace*{2cm}\times \big\langle\mc T H_\tn{T}(\tau_1) H_\tn{T}(\tau_2) H_\tn{T}(\tau_3)H_{\tn{T},-}^{R\sigma}\big\rangle_0~.
\end{split}\end{equation}
One can show that the imaginary part of the zero-coupling thermal expectation value $\langle\ldots H_{\tn{T},-}^{R\sigma}\rangle_0$ is non-vanishing only if it additionally involves each of the operators $H_{\tn{T},+}^{L\sigma}$, $H_{\tn{T},+}^{L\bar\sigma}$, $H_{\tn{T},-}^{R\bar\sigma}$ once; subsequently plugging in the lead contribution $\langle\mc T c_{s-k\downarrow}^\dagger(\tau) c_{sk\uparrow}^\dagger\rangle_0$ and carrying out the wide-band limit yields
\begin{equation}\label{eq:pt2}\begin{split}
&J  = -\frac{\Delta^2\Gamma^2\sin\phi}{2\pi^2}\sum_\sigma \int_0^{\beta}d\tau_1d\tau_2d\tau_3\int_{\mathbb{R}}d\epsilon_1 d\epsilon_2\\ & \times
\frac{f(\epsilon_1,\tau_3)f(\epsilon_2,\tau_1-\tau_2)}{E(\epsilon_1)E(\epsilon_2)}
\,\underbrace{\big\langle\mc T d_{1\bar\sigma}^\dagger(\tau_1)d_{1\sigma}^\dagger(\tau_2)d_{2\bar\sigma}^{\phantom{\dagger}}(\tau_3)d_{2\sigma}^{\phantom{\dagger}}\big\rangle_0}_{=g_\tn{dot}(\tau_1,\tau_2,\tau_3)}\,,
\end{split}\end{equation}
where $E(\epsilon)=\sqrt{\epsilon^2+\Delta^2}$, and 
\begin{equation}\begin{split}
f(\epsilon_k,\tau) & = -\frac{2E(\epsilon_k)e^{i\phi_s}}{\Delta} \big\langle\mc T c_{s-k\downarrow}^\dagger(\tau) c_{sk\uparrow}^\dagger \big\rangle_0 \\
& = e^{-E(\epsilon_k)|\tau|}-\frac{e^{E(\epsilon_k)\tau}+ e^{-E(\epsilon_k)\tau}}{1+e^{\beta E(\epsilon_k)}} \\[1ex]
& \approx e^{-E(\epsilon_k)|\tau|} - e^{-E(\epsilon_k)(\beta-|\tau|)}~.
\end{split}\end{equation}
We have simplified the last line assuming that $T\ll\Delta$ but emphasize that one cannot neglect $\exp(-E/T)$ even in the zero-temperature limit as the imaginary time integrations may give rise to a factor $\exp(E/T)$.

In order to evaluate the dot contribution $g_\tn{dot}(\tau_1,\tau_2,\tau_3)$, which is complicated by the lack of Wick's theorem, one first solves the many-particle eigenvalue problem $H_\tn{dot}|n\rangle = \lambda_n|n\rangle$. As we have assumed equal local- and nearest neighbor interactions, this can be achieved conveniently by rotating to a basis of bonding (b) and anti-bonding (a) states $d_{1,2\sigma}=(d_{b\sigma}\pm d_{a\sigma})/\sqrt{2}$. After inserting unit operators $1=\sum_n|n\rangle\langle n|$, one can then perform the $\tau$-integrals for a given time ordering and eventually set $T=0$. For $\tau_1>\tau_2>\tau_3$ we obtain (and likewise for the other five possibilities)
\begin{widetext}
\begin{equation}\label{eq:pt3}\begin{split}
\lim_{T\to0}& \int_0^\beta d\tau_3 \int_{\tau_3}^\beta d\tau_2\int_{\tau_2}^\beta d\tau_1 f(\epsilon_1,\tau_3)f(\epsilon_2,\tau_1-\tau_2)g_\tn{dot}(\tau_1,\tau_2,\tau_3) = \frac{1}{4}\sum_{npqz}
\langle z|d_{b\bar\sigma}^\dagger+d_{a\bar\sigma}^\dagger|n\rangle
\langle n |d_{b\sigma}^\dagger+d_{a\sigma}^\dagger|p\rangle \\
& \times \langle p |d_{b\bar\sigma}^{\phantom{\dagger}}-d_{a\bar\sigma}^{\phantom{\dagger}}|q\rangle
\langle q |d_{b\sigma}^{\phantom{\dagger}}-d_{a\sigma}^{\phantom{\dagger}}|z\rangle
\times \bigg\{ \frac{\delta_{\tilde\lambda_z,0}}{[E(\epsilon_1)+\tilde\lambda_q][E(\epsilon_2)+\tilde\lambda_n]\tilde\lambda_p}
+ \frac{\delta_{\tilde\lambda_p,0}}{[E(\epsilon_1)+\tilde\lambda_q][E(\epsilon_2)+\tilde\lambda_n]\tilde\lambda_z}\\
&~~ - \frac{\delta_{\tilde\lambda_q,0}}{[E(\epsilon_1)+\tilde\lambda_p][E(\epsilon_1)+\tilde\lambda_z][E(\epsilon_1)+E(\epsilon_2)+\tilde\lambda_n]}
- \frac{\delta_{\tilde\lambda_n,0}}{[E(\epsilon_2)+\tilde\lambda_p][E(\epsilon_2)+\tilde\lambda_z][E(\epsilon_1)+E(\epsilon_2)+\tilde\lambda_q]} \bigg\}~,
\end{split}\end{equation} 
\end{widetext}
with $\tilde\lambda_n=\lambda_n-\lambda_0$ denoting the difference to the smallest eigenvalue $\lambda_0$. Equation (\ref{eq:pt3}) illustrates that even in the zero-temperature limit it is in general not sufficient to consider only the corresponding ground state $|z\rangle=|0\rangle$ when computing the thermal expectation value $\langle\ldots\rangle_0$.\cite{commentpt} Namely, it turns out that if $|0\rangle$ appears as any of the intermediate states $|n,p,q\rangle$, a factor $\exp(\tilde\lambda_{n,q,p}/T)$ cancels the thermal weight $\exp(-\tilde\lambda_{n,q,p}/T)$. By combining Eq.~(\ref{eq:pt3}) as well as its five analogues with Eq.~(\ref{eq:pt2}), one can finally calculate the current by carrying out the remaining energy integrals numerically (or even analytically). More importantly, an intuitive understanding can be obtained by interpreting $J$ in terms of four successive tunnel events, each of which being described by one of the factors $\langle n|d^{(\dagger)}_{a,b\sigma}|q\rangle$.

\begin{figure*}[t]
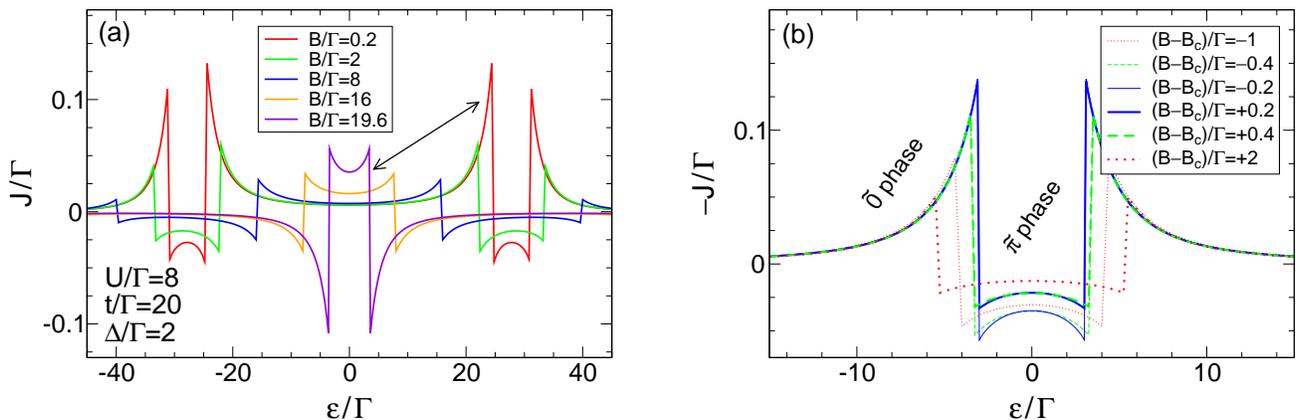

\centering
\includegraphics[width=0.45\linewidth,clip]{frg.eps}\hspace*{0.05\linewidth}
\includegraphics[width=0.45\linewidth,clip]{frg2.eps}
\caption {(Color online) Functional RG calculation for the gate-voltage $\epsilon$ dependence of the Josephson current $J$ flowing through the serial quantum dot for generic system parameters $U/\Gamma=8$ (Coulomb interaction in units of the lead hybridization strength), $\Delta/\Gamma=2$, $\phi/\pi=0.5$ (BCS gap and phase difference), $t/\Gamma=20$ (inter-dot coupling), zero temperature, and various Zeeman fields: (a) $B/\Gamma=0.2,2,8,16,19.6$ (from bottom to top at $\epsilon=0$) (b) $B/\Gamma=19,19.6,19.8,20.2,20.4,22$ or $(B-B_c)/\Gamma=-1,-0.4-0.2,0.2,0.4,2$ with $B_c=t$; see the main text for details.}
\label{fig:frg}
\end{figure*}

\section{Results}
\label{sec:results}

\subsection{Single-level case}
\label{sec:results.single}

In order to address the question in what respects the physics of our multi-level quantum dot in presence of a Zeeman field $B\approx t$ can be interpreted in a simple single-impurity fashion, it is helpful to give a short reminder of the latter. We focus on $\Delta=\infty$ which allows to compute both the current as well as the many-particle eigenenergies/states. If one thinks of the physics being governed by some interplay of Kondo screening and the formation of Cooper pairs -- and thus by the ratio $T_K/\Delta$ -- this large-gap limit might \textit{a priori} seem highly non-generic. However, various prior studies\cite{bcspaper,bcsdirekt,koenig2,koenig} showed that treating $\Delta=\infty$ is sufficient to qualitatively understand the single-level behavior (e.g., the parameter dependence of the phase boundary). In Section \ref{sec:results.atom}, we will illustrate that the same holds for our geometry.

The effective Hamiltonian associated with the single-level case at large gaps can be obtained in complete analogy with Eq.~(\ref{eq:hatom}):\cite{bcspaper,koenig2,koenig}
\begin{equation}\begin{split}
H_\tn{eff}^\tn{single} &= H_\tn{dot}^\tn{single} + 2 \Gamma\cos(\phi/2) \\
H_\tn{dot}^\tn{single} &= (\epsilon-U/2+B)n_\uparrow + (\epsilon-U/2-B)n_\downarrow + U n_\uparrow n_\downarrow~.
\end{split}\end{equation}
One can diagonalize it easily and eventually calculate the current from the phase derivative of the free energy [see Eq.~(\ref{eq:jatom})]. The results are shown in Fig.~\ref{fig:single}. At zero Zeeman field $B=0$, we find that either the ground state is a non-degenerate singlet with total spin $\{s=0,s_z=0\}$ and the first excited state is a twofold degenerate doublet $\{s=1/2,s_z=\pm 1/2\}$ (the $0$ phase with $J>0$), or vice versa (the $\pi$ phase where $J\leq0$; $J=0$ at $\Delta=\infty$). Decreasing $U$ (or $\phi$, or shifting $\epsilon$ away from half filling) favors the $0$ phase; at a critical $U_c$ (or $\phi_c$, or $\epsilon_c$), the system undergoes a first order level-crossing quantum phase transition as the nature of the ground states changes [see the $B=0$ curve $J(\epsilon)$ in Fig.~\ref{fig:single}(a); compare with Fig.~\ref{fig:single}(b) where $U<U_c$]. The current is discontinuous and flips its sign.

At finite $B\neq0$, the degeneracy of the states with $\{1/2,\pm1/2\}$ is lifted; but for small enough fields, the physics is still determined by an interplay of one non-degenerate and a pair of almost twofold degenerate states. As the singlet $\{0,0\}$ is not affected by the Zeeman field, the $\pi$ phase becomes larger with increasing $B$; one can moreover trigger a `Zeeman field $0$-$\pi$ transition' for $U<U_c$ by applying a sufficiently large $B$ [see Fig.~\ref{fig:single}(b)]. The effect of $B$ therefore \textit{seems} similar to that of $U$; and even the lineshape of $J(\epsilon,\phi)$ at large $B>0$ and $U=0$ heuristically resembles the one at $U>U_c$ and $B=0$. However, the absence of an (almost) twofold degenerate state in the former case renders it meaningless to speak about an interplay of $0$ and $\pi$ behavior.

\subsection{Functional RG for generic parameters}
\label{sec:results.frg}

We start our discussion of the serial quantum dot by calculating the zero-temperature Josephson current $J$ as a function of the gate voltage $\epsilon$ for experimentally-motivated parameters\cite{helene} of the Coulomb interaction ($U/\Gamma=8$), BCS gap ($\Delta/\Gamma=2$), and inter-dot coupling ($t/\Gamma=20$). To this end, we solve the functional renormalization group flow equations (\ref{eq:flowse}) and (\ref{eq:flowu}) numerically and subsequently compute $J$ from Eq.~(\ref{eq:j}). The results are shown in Fig.~\ref{fig:frg}. For small Zeeman fields, the lineshape of $J(\epsilon)$ features two copies of the single-level current gate-voltage relation (compare, e.g., with Ref.~\onlinecite{bcspaper} or with Fig.~\ref{fig:single}). This is easily understood from the fact that $H_\tn{dot}$ can be diagonalized by rotating to bonding and anti-bonding single-particle levels (see Sec.~\ref{sec:pt}); they are well-separated at $t\gg\Gamma$, and thus single-impurity behavior occurs if either is at the Fermi level ($\epsilon\approx\pm \epsilon_0$).

As $B$ increases, the size of the regions with $J<0$ increases; but again it eventually becomes meaningless to strictly speak about $0$ or $\pi$ regimes. On the other hand, one readily notices that at $B=B_c=t$:\cite{commentname} (i) The single-particle ($U=0$) energies of the anti-bonding spin-up and bonding spin-down states become equal;\cite{commentbc} they cross the Fermi level at $\epsilon=0$. (ii) The smallest eigenvalue of $H_\tn{dot}$ (at $U>0$) with two particles, which is the smallest overall one close to $\epsilon=0$, becomes twofold degenerate; the associated spin configuration is either a singlet or (one out of a) triplet. Both motivates us to investigate whether for $B\approx B_c$, $\epsilon\approx0$ the physics can again be described in a pure (and simple) single-impurity fashion. This turned out to be the case for normal leads where finite-$B$ Kondo ridges appear.\cite{schnuggi}

The Josephson current $J(\epsilon)$ of the multi-level quantum dot with $B\approx B_c$, $\epsilon\approx0$ indeed strikingly resembles the one for $B\approx0$, $\epsilon\approx\pm\epsilon_0$ up to an irrelevant overall sign [see Fig.~\ref{fig:frg} and compare the curves in (a) which are connected by the arrow; compare (b) with Ref.~\onlinecite{bcspaper} or with Fig.~\ref{fig:single}(a)]. Most importantly, there are discontinuities associated with a sign change; and we find that the regimes of negative and positive current -- from now on denoted $\tilde0$ and $\tilde\pi$, respectively -- feature precisely the same dependencies on system parameters as the $0$ and $\pi$ phases of a single impurity. Namely, decreasing $U$, $\epsilon$, $|B-B_c|$, or $\pi-\phi$ (an overall sign corresponds to a shift $\phi\to\phi+\pi$) favors the $\tilde0$ regime. Similar to the lineshape of $J(\epsilon)$, the current-phase relation $J(\phi)$ around $B\approx B_c$, $\epsilon\approx0$ is analogous to the single-level case: It is half-sinusoidal (sinusoidal) in the $\tilde0$ ($\tilde\pi$) regime.\cite{bcspaper} Those are all certainly rather heuristic arguments; but they can be fully supported by an analytic treatment of the large-gap limit (Sec.~\ref{sec:results.atom}) where one can access the exact many-particle eigenenergies/states.

On the other hand, there is one clear and interesting difference to the single-level case. Namely, for $U>U_c$ the current features another discontinuity at $B=B_c$ \textit{in addition} to the $\tilde0$-$\tilde\pi$ transition taking place at $U=U_c$ (or some $\epsilon=\epsilon_c$, or $\phi=\phi_c$). It is associated with a change in magnitude of $J$ but not a sign flip. This is illustrated in Fig.~\ref{fig:frg}(b); note that the curves for $(B-B_c)/\Gamma=-0.4,-0.2$ as well as for $(B-B_c)/\Gamma=0.2,0.4$ coincide, demonstrating that the current indeed exhibits a jump in the $\tilde\pi$ phase (a very detailed numerical investigation of this discontinuity seems superfluous as we can show analytically that it appears for $\Delta=\infty$; see Sec.~\ref{sec:results.atom}). On both sides -- i.e. for all $B\approx B_c$ -- of this supposedly first-order quantum phase transition, one observes $\tilde\pi$ phase behavior in the spirit outlined in the previous paragraph. Again, this is rather heuristic; and in order to obtain a more thorough understanding, we now analytically investigate the atomic limit ($\Delta=\infty$) and moreover carry out a perturbation expansion in the tunnel coupling $\Gamma$. We finally note that this altogether scenario (summarized by Fig.~\ref{fig:frg}) is generic -- it is not altered qualitatively if different local- and nearest neighbor interactions, level detunings or source-drain coupling asymmetries are introduced.

\begin{figure}[t]
\centering
\includegraphics[width=0.9\linewidth,clip]{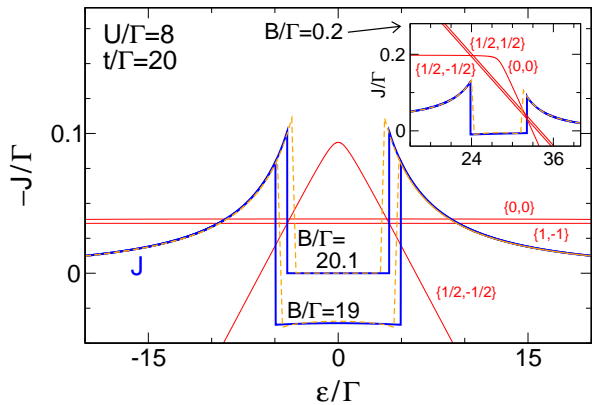}
\caption {(Color online) Exact results for the zero-temperature Josephson current in the infinite-gap limit (blue thick lines) compared to approximate FRG data obtained at $\Delta/\Gamma=2000$ (orange dashed lines) for three different Zeeman fields. The parameters are the same as in Fig.~\ref{fig:frg}. Thin red lines display the three lowest many-particle eigenenergies for $B/\Gamma=20.1$ (main panel) and $B/\Gamma=0.2$ (inset) in arbitrary units. They can be characterized by total spin quantum numbers $\{s,s_z\}$. For $B\approx B_c=t$, there is a level-crossing phase transition between a (nearly) twofold degenerate ground state formed by $\{0,0\}$ and $\{1,-1\}$ and a non-degenerate one as $\epsilon$ is shifted away from the Fermi energy. This is the typical behavior of a single impurity at small Zeeman fields [which features an almost identical lineshape $J(\epsilon)$; see the inset as well as Fig.~\ref{fig:single}(a)]. In contrast to the latter, however, the magnitude of the current is different in each of the states $\{0,0\}$ and $\{1,-1\}$; there is another level-crossing phase transition at $B=B_c$ where the ground state changes from $\{0,0\}$ to $\{1,-1\}$ or vice versa (this can be shown analytically; see also Fig.~\ref{fig:atom2}).}
\label{fig:atom1}
\end{figure}

\begin{figure}[t]
\centering
\includegraphics[width=0.9\linewidth,clip]{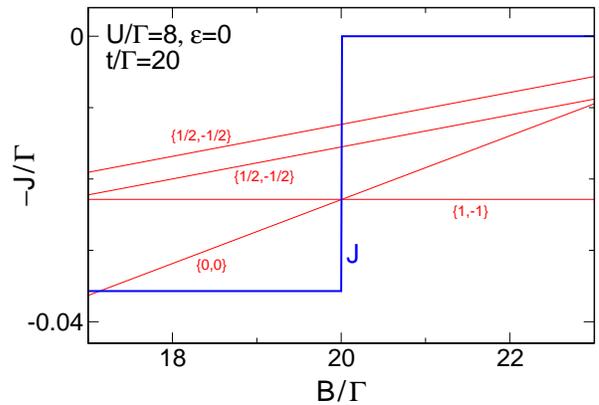}
\caption {(Color online) Atomic-limit ($\Delta=\infty$) Josephson current (thick line) and four lowest eigenenergies (thin lines; arbitrary units) as a function of the Zeeman field $B$ at $\epsilon=0$ (other parameters are as in Fig.~\ref{fig:frg}). Since $U>U_c$, the ground state is almost twofold degenerate for $B\approx B_c$; the involved states have a total spin $\{0,0\}$ or $\{1,-1\}$, respectively. Thus, the spin configuration is a singlet for $B<B_c$ but (one out of a) triplet for $B>B_c$ -- an additional level-crossing phase transition occurs at $B=B_c$ in the $\tilde\pi$ regime; the current is discontinuous.}
\label{fig:atom2}
\end{figure}

\subsection{Insights from the atomic limit}
\label{sec:results.atom}

As outlined in Section \ref{sec:atom}, our multi-level quantum dot geometry can be treated analytically in the large-gap limit even in presence of two-particle interactions; we merely need to diagonalize the effective Hamiltonian of Eq.~(\ref{eq:hatom}) in the underlying 16-dimensional Hilbert space and subsequently compute $J$ from the phase derivative of the free energy [see Eq.~(\ref{eq:jatom})]. The current at $\Delta=\infty$ is similar to the one at $\Delta/\Gamma=2$ (see Fig.~\ref{fig:atom1} and compare with Fig.~\ref{fig:frg}); this is consistent with prior studies of the single-level case.\cite{bcspaper,bcsdirekt,koenig2,koenig} In particular, its lineshape features single-impurity characteristics for $B\approx B_c=t$, $\epsilon\approx0$ accompanied by an additional discontinuity at $B=B_c$ in the $\tilde\pi$ regime. Again, $J$ is larger at $B<B_c$ and even vanishes for $B>B_c$ at $\Delta=\infty$. Before trying to understand this more thoroughly, let us just note that the results at $\Delta=\infty$ can be employed to benchmark our FRG approach. It turns out that despite the fact that the FRG embodies an approximation which can \textit{a priori} be justified only for small values of the Coulomb interaction, its results compare decently with the exact reference even for fairly large $U/\Gamma=8$ (compare thick solid and dashed lines in Fig.~\ref{fig:atom1}). This observation is again in agreement with the single-level case.\cite{bcspaper,bcsdirekt} 

Most importantly (and frequently referred to above), one can try to learn more about the physics by investigating the many-particle eigenvalues and eigenstates of $H_\tn{eff}$ and by characterizing them in terms of the square and $z$-component of the total spin $\vec s$. For small $B$ as well as $\epsilon$ chosen such that either the bonding or anti-bonding level is close to the Fermi energy, one recovers (see the inset to Fig.~\ref{fig:atom1}) the scenario outlined in Sec.~\ref{sec:results.single}: The ground state is either a non-degenerate spin singlet with $\{s=0,s_z=0\}$ (the $0$ phase) or an almost twofold degenerate state associated with a free spin $\{1/2,\pm 1/2\}$ (the $\pi$ phase; the ground state is exactly twofold degenerate at $B=0$). However, the very same degeneracy properties appear for $B\approx t$:\cite{degeneracy} What we labeled $\tilde0$ and $\tilde\pi$ regime is always associated with a non-degenerate and a nearly (exactly at $B=B_c=t$) doubly degenerate state, respectively. E.g., at $U>U_c$ the ground state is nearly twofold degenerate around $\epsilon\approx0$ but becomes non-degenerate as the gate voltage increases beyond $\pm\epsilon_c$ (see the thin lines in the main panel of Fig.~\ref{fig:atom1} and compare with the inset). At small $U<U_c$, the ground state is always non-degenerate whereas the first excited state is nearly twofold degenerate, and the current is continuous. Summarizing, this interplay of a non-degenerate and a (nearly) twofold degenerate state supports our conjecture of a level-crossing $\tilde0$-$\tilde\pi$ transition governing our multi-level quantum dot for Zeeman fields $B\approx B_c$.

The non-degenerate and pair of degenerate states involved are characterized by total spin quantum numbers $\{s=1/2,s_z=-1/2\}$ ($\tilde0$ phase) as well as $\{0,0\}$, $\{1,-1\}$ ($\tilde\pi$ phase), respectively.\cite{degeneracy,directionb} In the $\tilde\pi$ regime, the ground state is thus exactly twofold degenerate only at $B=B_c=t$; for small deviations $B\approx B_c$, the degeneracy is lifted. The same holds in the single-level case; but there the current is obviously identical for both $\{1/2,\pm1/2\}$ (i.e., independent of the sign of $B$). For the problem at hand, however, the ground state spin configuration is different for $B-B_c<0$ and $B-B_c>0$; it is a singlet in the former but (one out of a) a triplet in the latter case. This renders it reasonable that also $J$ is different, and thus another level crossing phase transition occurs at $B=B_c$ in the $\tilde\pi$ regime (see Fig.~\ref{fig:atom2}). Even though it might be intuitively clear that the current is smaller in the $\{1,-1\}$ configuration\cite{koenig2} -- a triplet should prevent Cooper pair tunneling -- we will now substantiate that by a perturbation theory approach.

\begin{figure}[t]
\centering
\includegraphics[width=0.9\linewidth,clip]{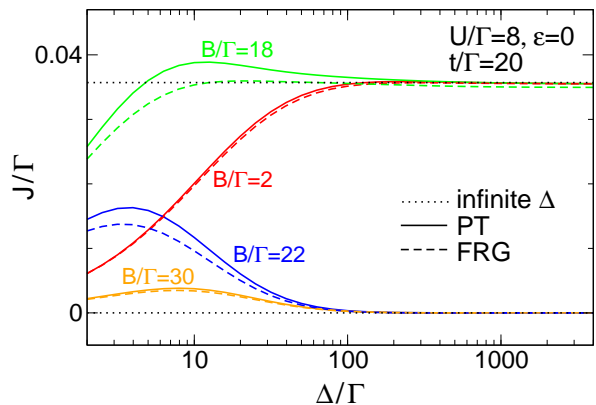}
\caption {(Color online) Zero-temperature Josephson current as a function of the BCS gap $\Delta$ for various Zeeman fields (all other parameters are as in Fig.~\ref{fig:frg}). Solid lines were obtained by carrying out perturbation theory in the lead coupling $\Gamma$, dashed lines display FRG data. Dotted lines show the atomic-limit result. Remember that the current in the $\tilde\pi$ phase vanishes at $\Delta\to\infty$ for $B>B_c$ but remains finite for $B<B_c$.}
\label{fig:pt}
\end{figure}

\subsection{Insights from perturbation theory}
\label{sec:results.pt}

The expansion of the Josephson current to fourth order in the level-lead coupling Hamiltonian $H_\tn{T}$ is given by Eq.~(\ref{eq:pt2}). The imaginary time integrals can be carried out analytically [see Eq.~(\ref{eq:pt3})], but in general one needs to resort to numerics to perform the remaining two energy integrations. The resulting zero-temperature current as a function of the gap is shown in Fig.~\ref{fig:pt}. One can easily understand that it again features a discontinuity around $B=B_c$, $\epsilon=0$; this is merely due to the fact that the smallest eigenvalue of $H_\tn{dot}$ corresponds to a spin singlet state $d_{b\uparrow}^\dagger d_{b\downarrow}^\dagger|\tn{vac}\rangle$ for $B<B_c$ but a triplet configuration $d_{a\downarrow}^\dagger d_{b\downarrow}^\dagger|\tn{vac}\rangle$ for $B>B_c$,\cite{directionb} and thus different terms contribute to Eq.~(\ref{eq:pt3}) in each case. More generally, one observes a decent quantitative agreement with FRG data even at $\Delta/\Gamma=2$ despite the fact that the expansion w.r.t.~$H_\tn{T}$ can \textit{a priori} be justified only for large gaps.

\begin{figure}[b]
\centering
\includegraphics[width=0.6\linewidth,clip]{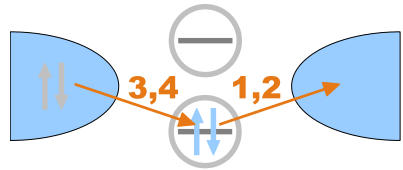}
\caption {(Color online) The current through a singlet dot configuration interpreted in terms of four (1,2,3,4) successive tunnel events in the basis of bonding and anti-bonding molecular levels. In the depicted contribution, only two intermediate states contain an unpaired lead electron (the one between 2 and 3 does not). The latter require a quasiparticle excitation energy $E(\epsilon_i)=\sqrt{\epsilon_i^2+\Delta^2}$ and are thus unfavorable at large gaps. If the dot is initially in a triplet configuration $d_{a\downarrow}^\dagger d_{b\downarrow}^\dagger|\tn{vac}\rangle$, no such process exists, and the current is suppressed more strongly.}
\label{fig:pt2}
\end{figure}

Having established that perturbation theory gives sensible results, we now additionally turn to the limit of $\Delta\to\infty$; this large-gap case is again particularly simple since only a few contributions to Eq.~(\ref{eq:pt3}) as well as to its five analogues originating from different time orderings remain finite for $B<B_c$ (and the current vanishes for $B>B_c$). The $\epsilon_i$-integrals in Eq.~(\ref{eq:pt2}) can then be carried out analytically:
\begin{equation}
J \stackrel{\Delta\to\infty}{=}
\begin{cases}
\frac{\Gamma^2\sin\phi}{2}\left(\frac{1}{U+t+\epsilon} + \frac{1}{U+t-\epsilon} \right) & B<B_c \\
0 & B> B_c~.
\end{cases}
\end{equation}
For $B<B_c$, one can moreover intuitively interpret every (of the few) non-vanishing terms in Eq.~(\ref{eq:pt3}) by associating a single tunnel event with each of the four factors $\langle n|d^{(\dagger)}_{a,b\sigma}|q\rangle$. Fortunately, there are only two qualitatively different scenarios: The two electrons occupying the dot\cite{commentpt2} are first transferred to the right lead, and then the original dot configuration is restored by two electrons tunneling in from the left (see Fig.~\ref{fig:pt2}); or alternatively, an additional Cooper pair from the left hops onto the dot and then further on to the right. In either case, there are two intermediate configurations that contain a single unpaired lead electron. Both give rise to a (quasiparticle) energy denominator $E(\epsilon_i)=\sqrt{\epsilon_i^2+\Delta^2}$, but the current remains finite in the $\Delta\to\infty$ limit due to the explicit prefactor $\Delta^2$ in Eq.~(\ref{eq:pt2}). On the other hand, it is quite clear that no similar process exists for $B>B_c$ where the dot is initially in a triplet state $d_{a\downarrow}^\dagger d_{b\downarrow}^\dagger|\tn{vac}\rangle$ (i.e., occupied by two electrons with equal spin); but since there is no possibility for an intermediate state that does not exhibit a lead quasiparticle excitation, the current is suppressed more strongly at large gaps. This eventually provides a simple cartoon picture for the discontinuity at $B=B_c$: A Cooper pair can tunnel more easily through a singlet than a triplet dot configuration.

\section{Conclusion}

We have studied the equilibrium supercurrent $J$ through a multi-level quantum dot in the molecular regime of large couplings $t$ between the impurities. If a Zeeman field $B\approx t$ is introduced such that the lowest two-particle energies (whose total spin configuration is either $\{s=0,s_z=0\}$ or $\{s=1,s_z=-1\}$) of the isolated dot are equal, several characteristics of the system -- such as the appearance of a $\tilde0$-$\tilde\pi$ transition as well as the corresponding lineshapes and parameter dependencies -- are similar to the well-known case of a single impurity. On the other hand, the current features an additional discontinuity at $B=B_c=t$ in the $\tilde\pi$ regime; it can be attributed to the fact that a Cooper pair can tunnel more easily through a spin singlet state than through a triplet one. This scenario should be generic for any more complex quantum dot geometry close a singlet-triplet degeneracy.

\section*{Acknowledgments}
We thank S.~Droste and J.~Splettst\"o\ss er for useful discussions and are grateful to the Deutsche Forschungsgemeinschaft for support via FOR723 as well as KA3360-1/1 (C.K.).

\appendix

\begin{figure}[t]
\centering
\includegraphics[width=0.9\linewidth,clip]{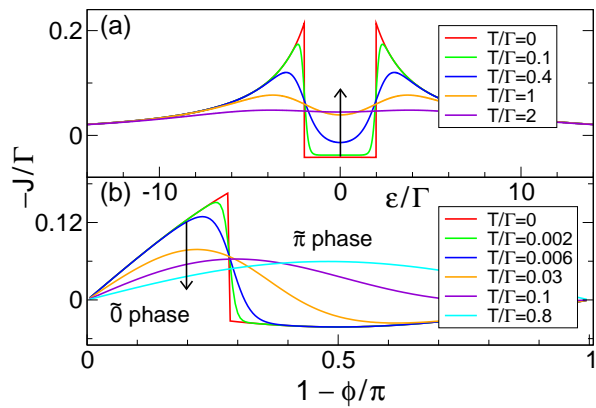}
\caption {(Color online) Atomic-limit calculation of the Josephson current as a function of (a) the gate voltage $\epsilon$ at $\phi/\pi=0.5$ and (b) the phase difference $\phi$ at $\epsilon/\Gamma=1.9$ for $U/\Gamma=4$, $t/\Gamma=20$, $B/\Gamma=19.9$ (`close to singlet-triplet degeneracy') and various temperatures $T$ ($J$ varies with increasing $T$ as idicated by the arrows). Note that this is an exact result. }
\label{fig:atom3}
\end{figure}

\section{Finite temperatures}

The analytic treatment of the atomic limit outlined in Sec.~\ref{sec:atom} allows to readily compute $J$ not only at $T=0$ but also for finite temperatures. This was not the main focus in the single-level case where the current at $T=0$ and $\Delta=\infty$ always vanishes in the $\pi$ regime [see Fig.~\ref{fig:single}(a)]; the discussion of the large-gap limit therefore mostly concentrated on the parameter dependence of the phase boundary. For the problem at hand, $J$ is non-zero in the $\tilde\pi$ phase for Zeeman fields $B<B_c$ even at $\Delta=\infty$, and this motivates us to briefly present (exact!) results for the current at finite temperatures (see Fig.~\ref{fig:atom3}). Not unexpected, the discontinuities in $J$ are smeared at $T>0$; this is similar to the single-level case where non-zero $T>0$ were studied at finite gaps by virtue of the numerical renormalization group\cite{choi,bcspaper} or quantum Monte Carlo.\cite{se,qmc} Moreover, the current-phase relation -- which is generally (half-) sinusoidal in the $\tilde\pi$ ($\tilde0$) regimes -- becomes completely sinusoidal at sufficiently large temperatures. This is once again completely analogous to the single-level case.



\begin{thebibliography}{*}
\bibitem{absfootnote}
E.g., the results of a recent experiment (Ref.~\onlinecite{absexp}) directly detecting the Andreev bound states (which carry an important part of the Josephson current) were interpreted in the light of a multi-level/multi-dot geometry.
\bibitem{directionb}
Without loss of generality, we focus exclusively on positive Zeeman fields $B>0$.
\bibitem{commentpt}
This was already pointed out for the single-level case\cite{novotny} but might have been overlooked in a previous discussion of the singlet-triplet transition for a general two-level dot.\cite{lee}
\bibitem{commentname}
The labeling of $B_c$ intentionally indicates `critical' -- the reason for this will become clear at the end of Sec.~\ref{sec:results.atom}.
\bibitem{commentbc}
Both the hopping $t$ and the field $B$ certainly get renormalized by the Coulomb interaction. For equal local- and nearest neighbor interactions, however, we observe within our FRG approach that the two effects cancel each other (in agreement with the case of normal leads discussed in Ref.~\onlinecite{schnuggi}), motivating $B_c=t$.
\bibitem{degeneracy}
Since the current associated with the full Hamiltonian $H$ at $\Delta=\infty$ and that of the effective $H_\tn{eff}$ is the same, the properties of the eigenvalues/states of $H_\tn{eff}$ are to be interpreted as actually mimicking those of $H$. This is not to be confused with our discussion of the many-particle eigenvalues of $H_\tn{dot}$; the latter is supposed to merely motivate looking for simple physics around $B_c=t$.
\bibitem{commentpt2}
As discussed in Sec.~\ref{sec:results.pt}, in the perturbation theory approach it is not sufficient to consider only the ground state when computing the thermal expectation value of the current at zero temperature. However, it turns out that the only terms that may contribute at $T=0$ are those where the ground state appears as one of the states $|n,p,q,z\rangle$ in Eq.~(\ref{eq:pt3}). By reordering the four factors, one can thus always think of starting in a ground state configuration when qualitatively interpreting the current in terms of successive tunnel events.
\end{thebibliography}
\end{document}